\documentclass{ws-ijmpa}
\usepackage[super,compress]{cite}
\usepackage{graphicx}
\usepackage{cite}
\begin{document}
\markboth{William H. Pannell}{The Intersection between Dual Potential and SL(2) Algebraic Spectral Problems}

\catchline{}{}{}{}{}
%
%%%%%%%%%%%%%%%%%%%%%%%%%%%%%%%%%%%%%%%%%%%%%%%%%%%%%%%%%%%%%%%%%%%%

\title{The Intersection between Dual Potential and SL(2) Algebraic Spectral Problems}

\author{William H. Pannell}

\address{School of Physics and Astronomy, University of Minnesota Twin-Cities, 116 Church St SE\\
Minneapolis, Minnesota 55455, United States\\
panne022@umn.edu}

\maketitle

%\begin{history}
%\received{}
%\revised{}
%\end{history}

\begin{abstract}
The relation between certain Hamiltonians, known as dual, or partner Hamiltonians, under the transformation $x{\rightarrow}\bar{x}^{\bar{\alpha}}$ has long been used as a method of simplifying spectral problems in quantum mechanics. This paper seeks to examine this further by expressing such Hamiltonians in terms of the generators of SL(2) algebra, which provides another method of solving spectral problems. It appears that doing so greatly restricts the set of allowable potentials, with the only non-trivial potentials allowed being the Coulomb $\frac{1}{r}$ potential and the Harmonic Oscillator $r^2$ potential, for both of which the SL(2) expression is already known. It also appears that, by utilizing both the partner potential transformation and the formalism of the Lie-algebraic construction of quantum mechanics, it may be possible to construct part of a Hamiltonian's spectrum from the quasi-solvability of its partner Hamiltonian.

\end{abstract}

%\ccode{PACS numbers:}

%\tableofcontents

\section{\label{sec:level1}Introduction}
 A famous classical relationship relates the Hamiltonians governing the $\frac{1}{r}$ potential and the $r^2$ potential. This pairing between the oscillator and Coulomb potential has been known since the time of Newton, who used it to solve problems pertaining to planetary motion\cite{newton}. This relationship survived the passage to quantum mechanics, where the transformation now applies to the radial part of the Schrödinger equation\cite{rowley}.
 
This relationship is part of a wider framework\cite{gazeau}, in which any given potential of the form $V(x){\propto}x^\alpha$ can be related to another potential of the nearly identical form $V(\bar{x}){\propto}\bar{x}^{\bar{\alpha}}$\cite{bars2020}, where the powers of the potentials are governed by the relation \begin{equation}
    (\alpha+2)(\bar{\alpha}+2)=4.
\end{equation}
Such potentials are known as partner, or dual potentials, and are so closely related that by taking the substitution \begin{equation}
    x=\bar{x}^{-\frac{\bar{\alpha}}{\alpha}}
\end{equation}
the Hamiltonian of one dual potential transforms exactly into the Hamiltonian of the other\cite{johnson}.

When the coordinate transformation is used in one Hamiltonian, and the appropriate quasi-gauge is selected to remove all of the resulting first derivative terms, it becomes exactly its partner Hamiltonian\cite{ahmed}, thus requiring that both the eigenfunctions and spectra of the respective Hamiltonians be related. Given these relations, it is then possible to use the transformation between Hamiltonians to simplify the problem of determining their respective spectra, easily solving systems in which one of the partners is already readily solvable.

Alternatively, it is possible to determine the spectra of a given Hamiltonian by using algebraic methods. This property certain quantum mechanical systems has long been known, with perhaps the most famous example being the expression of the harmonic oscillator Hamiltonian in terms of raising and lowering operators. In such problems, the solution for the spectra could be expressed purely in terms of the algebra, and was thus completely solvable. In the 1980s it was discovered that it was possible to extend this system of algebraization to a further group of Hamiltonians, which did not necessarilly satisfy the requirements of exact solvability\cite{turbiner}, and was later extended into the Lie-algebraic construction of quantum mechanics\cite{kamran}. An alternative representation, known as the analytic approach was then developed by Ushveridze\cite{ushveridze}. For such Hamiltonians, it was possible to use the SL(2) algebra to express either part of the spectra, or in certain cases the entirety of the spectrum, so that the system became quasi-solvable. 

The generators of SL(2) are given by
\begin{equation}
\begin{split}
    T^+=2j\xi-\xi^2\frac{d}{d\xi}, \\
    T^0=-j+\xi\frac{d}{d\xi}, \\
    T^-=\frac{d}{d\xi}.
    \label{eq:five}
\end{split}
\end{equation}
The above representation is only useful when $j$ is a semi-integer, for that is the case in which the representations become finite-dimensional\cite{shif98}. For such systems, $j$ can then be used to represent spin, with $2j+1$ levels corresponding to the $2s+1$ spin levels. These generators act on polynomials of $\xi$, and therefore, as $\xi$ is a function of the coordinate $x$, on wavefunctions $\psi(x)$. 

As the Hamiltonian will not include terms with third order derivatives of $x$, and as any triple combination of generators will necessarily include such a derivative term, a given Hamiltonian $H$ must be expressible as
\begin{equation}
    H={\sum}C_{ab}T^aT^b+{\sum}C_aT^a,
\end{equation}
where the constants $C_{ab}$ and $C_a$ are determined by the parameters present within the Hamiltonian. Given the form of the generators from Eq.~(\ref{eq:five}), this sum can expanded as
\begin{equation}
    H=-\frac{1}{2}P_4(\xi)\frac{d^2}{d\xi^2}+P_3(\xi)\frac{d}{d\xi}+P_2(\xi),
    \label{eq:six}
\end{equation}
where the $P_4(\xi)$, $P_3(\xi)$, and $P_2(\xi)$ are polynomials in $\xi$ whose form depends upon the generators taken to represent $H$\cite{shif98}. The true Hamiltonian $H(x)$ can then be recovered from this polynomial expression by defining the relation between $x$ and $\xi$ to be
\begin{equation}
    x={\int}d{\xi}P_4(\xi)^{-1/2}.
    \label{eq:eight}
\end{equation}
With this definition, Eq.~(\ref{eq:six}) becomes $$H=-\frac{1}{2}\frac{d^2}{dx^2}+A(x)\frac{d}{dx}+{\Delta}V.$$
Here, the quasi-gauge $A(x)$ and the quasi-gauge potential ${\Delta}V$ are defined as
\begin{equation}
\begin{split} 
A(x)=\frac{P_4'/4+P_3}{P_4^{1/2}}\Bigr|_{\substack{\xi=\xi(x)}}, \\
{\Delta}V=P_2(\xi)\Bigr|_{\substack{\xi=\xi(x)}}, \\
V(x)={\Delta}V+\frac{1}{2}A^2(x)-\frac{1}{2}A'(x),
\label{eq:nine}
\end{split}
\end{equation}
where the prime in $P_4'$ represents a derivative with respect to $\xi$, while the prime in $A'(x)$ represents a derivative with respect to $x$. Defining $$a(x)={\int}Adx,$$ $$\psi{\rightarrow}e^{-a(x)}\psi,$$ removes the terms in $H$ proportional to the first derivative of $x$, resulting in the proper Hamiltonian.

This paper examines the relationship between Hamiltonians which are expressible in terms of the SL(2) algebra, and those which are able to utilize the partner potential transformation. Section II covers how the restriction that both partner potential Hamiltonians be expressible in terms of the generators of SL(2) impacts the set of allowable potentials. Section III then covers the specific example of the partner potential transformation between the Coulomb and Harmonic Oscillator potentials using the formalism of SL(2) algebra, and examines the relationship between the spectra of the Hamiltonians. Appendix A covers the problem of the Coulomb and Harmonic Oscillator potentials without the use of SL(2), and derives the proportionality between the partner Hamiltonians.

\section{\label{sec:level3}SL(2) Restrictions on $\alpha$ and $\bar{\alpha}$}

Suppose that $x=\xi^k$ for some $k{\in}\mathbb{R}$. Then as a consequence it must hold that $$\frac{d^2}{dx^2}=\frac{1}{k^2}\xi^{2-2k}\frac{d^2}{d\xi^2}+\frac{1-k}{k^2}\xi^{1-2k}\frac{d}{d\xi}.$$
In order for $\frac{d^2}{dx^2}$ to be in the SL(2) algebra, it must be that $2-k=n$ for some $n\in\mathbb{Z}$ such that $0{\leq}n{\leq}4$. Then, $x^2=\xi^{2k}=\xi^{2-n}=\xi^{\gamma}$ for some $|\gamma|\leq2$, so that both $x$ and $x^2$ must be small powers of $\xi$. It then must also be the case that $\bar{x}^2=\bar{\xi}^{\bar{\gamma}}$ for $|\bar{\gamma}|\leq2$.

This restriction requires that $$|k|\leq1,$$ $$|\bar{k}|\leq1,$$ $$\frac{d}{d\xi}=kx^{1-1/k}\frac{d}{dx}.$$ In order to construct the $V(x)=x^\alpha$ term in $H_G$, one must add either a $T^+$ or a $T^-$ term. Then, ${\Delta}V$, $A'(x)$, and $A(x)^2$ are proportional to powers of x that are greater than or equal to $-2$, so that this formulation based on the SL(2) algebra can only account for potentials with $x^\alpha$ such that
\begin{equation}
\alpha\geq-2.
\end{equation}
As the same restrictions apply to $\bar{k}$ and $\bar{x}$, it must also be true that
\begin{equation}
    \bar{\alpha}\geq-2.
\end{equation}
This is the first restriction on the allowable dual potentials, and appears to be derived from the requirement that both $x$ and $x^2$ be small powers of $\xi$.

As partner potentials are given by $V(x)={\lambda}x^\alpha$, the form of $V(x)$ in Eq.~(\ref{eq:nine}) indicates that ${\int}d{\xi}P_4(\xi)^{-1/2}$ is equal to some polynomial in $\xi$. In order to ensure this, it must be the case that only one term in $P_4(\xi)$ is non-zero, so that $$P_4(\xi){\propto}\xi^n$$
for $n=0,1,3,4$. The $\xi^2$ term in $P_4$ cannot be used, as this leads to a $\xi=e^x$ term, which cannot account for a potential of the form necessary for partner potentials. This choice of $P_4$ restricts $x$ to be related to $\xi$ by \begin{equation}
    x=c\xi^{\frac{2-n}{2}}
\end{equation} for a constant of proportionality $c$. Selecting the $\xi^n$ term in $P_4$ additionally requires that $P_3$ and $P_2$ include $\xi^{n-1}$ and $\xi^{n-2}$ terms, as long as the exponents are positive. Then, given the relationship between $x$ and $\xi$, $$P_2{\propto}x^{-2},$$ $$A{\propto}\xi^{n-1-n/2}=x^{-1}$$ so that the potential term (Eq.~(\ref{eq:nine})) due to this choice of the polynomials is $$V(x){\propto}x^{-2}.$$ As $\alpha=-2$ is not an allowable partner potential, an additional term must be added to $P_3$, so that $$P_3{\propto}\xi^{n-1}+\xi^\gamma$$ for $\gamma\leq2$. This additional term will also add a term to $P_2$, so that $$P_2{\propto}\xi^{n-2}+\xi^{\gamma-1}$$ provided that $\gamma-1$ is non-negative. Then, $${\Delta}V{\propto}x^{-2}+x^{\frac{2\beta_2}{2-n}},$$ 
$$A(x){\propto}x^{-1}+x^{\frac{2\beta_3-n}{2-n}},$$ so that 
$$A^2{\propto}x^{-2}+x^{\frac{4\beta_3-2n}{2-n}}+x^{\frac{2\beta_3-2}{2-n}},$$ $$A'{\propto}x^{-2}+x^{\frac{2\beta_3-2}{2-n}}.$$ 
As $n\in\mathbb{N}$ such that $n\leq4$, $|2-n|\leq2$, so that, as $\beta_3\leq2$, $\frac{\beta_3}{2-n}\in\mathbb{Z}$. Similarly, the other powers of $x$ within $A^2$ and $A'$ must be integers, so that it must hold that
\begin{equation}
    \alpha\in\mathbb{Z}.
\end{equation}
As the same restrictions apply equally to $\bar{x}$ as to $x$, it must also be true that
\begin{equation}
    \bar{\alpha}\in\mathbb{Z}.
\end{equation}

The restriction of $\alpha$ and $\bar{\alpha}$ to only be integers has the effect of reducing the number of allowable partner potentials down to six. To see this, one must examine how the requirement for integer $\alpha$ and $\bar{\alpha}$ impacts the dual potentials. As the dual potentials are given by $$(\alpha+2)(\bar{\alpha}+2)=4,$$ it is possible to write $\bar{\alpha}$ as \begin{equation}
    \bar{\alpha}=-\frac{2\alpha}{\alpha+2}.
\end{equation}
If $\alpha$ and $\bar{\alpha}$ are given such that $\alpha,\bar{\alpha}\in\mathbb{Z}$, then it must be possible to write $n=-2\alpha$ and $m=\alpha+2$ so that $\bar{\alpha}=\frac{n}{m}$. There are then two possibilities, for the partner potentials are split into a group of potentials that are both less than $-2$, and a group of potentials that are both greater than $-2$. 

If $\alpha\geq-2$, then $m\geq1$. As $m+\frac{n}{2}=2$ and as $n=\bar{\alpha}m$, $$m=\frac{2}{1+\frac{\bar{\alpha}}{2}}\geq1$$ so that $\bar{\alpha}\leq2$. Then there are only three possible integer potentials: $-1, 0,$ and $2$.

If $\alpha\leq-2$, then $m\leq-1$ so that $\bar{\alpha}\geq-6$. Then, again, there are only three possible integer partner potentials: $-6, -4,$ and $-3$. Thus, the only partner potentials that see both $\alpha\in\mathbb{Z}$ and $\bar{\alpha}\in\mathbb{Z}$ are those with one of
\begin{equation}
    \alpha=-6, -4, -3, -1, 0, 2
\end{equation}
Clearly, this restriction, combined with the restriction that $\alpha$, $\bar{\alpha}\geq-2$ requires that $\alpha$, $\bar{\alpha}=-1, 0,$ or $2$. Thus, the only potentials that have Hamiltonians solvable both by the methods of SL(2) and the dual transformation are those of the free particle, the harmonic oscillator, and the Coulomb potential.

\section{\label{sec:level4}A SL(2) Solution for $\frac{1}{r}$ and $r^2$ Potentials}
It is possible to construct a number of different representations of the $\frac{1}{r}$ and $r^2$ Hamiltonians within the confines of SL(2) Algebra. Each of the possible representations can be shown to transform properly under the dual correspondence, for under the substitution $x=\xi^{\frac{2-n}{2}}$ they all properly reduce to the Hamiltonians examined in Appendix A.
 
 \subsection{The Form of the Solution}
 This example considers the situation in which $\alpha=-1$, $\bar{\alpha}=2$, $n=4$, $\bar{n}=1$, so that $x=\xi^{-1}$, and $\bar{x}=\bar{\xi}^{\frac{1}{2}}$, where the Hamiltonians are related by Eq.~(\ref{eq:seven}). The Hamiltonian is then given by Eq.~(\ref{eq:six}), where the generator polynomials are $$P_4(\xi)=-2C_{++}\xi^4,$$ $$P_3(\xi)=-2kC_{++}\xi^3-C_+\xi^2,$$ $$P_2(\xi)=2jkC_{++}\xi^2+2jC_+\xi,$$ $$\bar{P}_4(\bar{\xi})=-4\bar{C}_{-0}\bar{\xi},$$ $$\bar{P}_3(\bar{\xi})=-\bar{C}_+\bar{\xi}^2-k\bar{C}_{-0},$$  $$\bar{P}_2(\bar{\xi})=2j\bar{C}_+\bar{\xi}.$$ As $\xi=\bar{\xi}^{-1}$, derivatives with respect to $\xi$ and $\bar{\xi}$ must be related by $$\frac{d}{d\xi}=-\bar{\xi}^2\frac{d}{d\bar{\xi}},$$ $$\frac{d^2}{d\xi^2}=\bar{\xi}^4\frac{d^2}{d\bar{\xi}^2}+2\bar{\xi}^3\frac{d}{d\bar{\xi}}.$$ As $P_4$ is related to the $\frac{d^2}{dx^2}$ term in $H$ rather than being part of the quasi-gauge that fixes $V(x)$, it should hold that $$P_4\frac{d^2}{d\xi^2}{\propto}\bar{x}^{2+2\frac{\bar{\alpha}}{\alpha}}\bar{P}_4\frac{d^2}{d\bar{\xi}^2}.$$
Given the form of $P_4$ and $\frac{d^2}{d\xi^2}$ from above, $$P_4\frac{d^2}{d\xi^2}=\bar{\xi}^{-4}(\bar{\xi}^4\frac{d^2}{d\bar{\xi}^2}+2\bar{\xi}^3\frac{d}{d\bar{\xi}})=\frac{d^2}{d\bar{\xi}^2}+2\bar{\xi}\frac{d}{d\bar{\xi}}.$$
Multiplying this by $\bar{x}^{-2-2\frac{\bar{\alpha}}{\alpha}}=\bar{\xi}$, the resulting equation is
\begin{equation}
   \bar{x}^{-2-2\frac{\bar{\alpha}}{\alpha}}P_4\frac{d^2}{d\xi^2}=\bar{\xi}\frac{d^2}{d\bar{\xi}^2}+2\bar{\xi}^2\frac{d}{d\bar{\xi}}.
\end{equation}
Clearly the first term is exactly $\bar{P}_4\frac{d^2}{d\bar{\xi}^2}$, while the second term is related to the transformation $u(r)=\bar{r}^{-\frac{1}{2}}\bar{u}(\bar{r})$, so that $P_4$ transforms correctly into part of $\bar{H}$.

Then, the polynomials $P_3$ and $P_2$ must transform as parts of the quasi-gauge potential given in Eq.~(\ref{eq:nine}). As it is $V(x)$, rather than any of its individual components, that transforms into $\bar{V}(\bar{x})$, it must hold that $P_3(\xi)$ and $P_2(\xi)$ transform into $\bar{P}_3(\bar{\xi})$ and $\bar{P}_2(\bar{\xi})$ according to
\begin{equation}
    [{\Delta}V+\frac{1}{2}A^2-\frac{1}{2}A'+E]=\frac{\alpha^2}{\bar{\alpha}^2}\bar{x}^{2+2\frac{\bar{\alpha}}{\alpha}}[{\Delta}\bar{V}+\frac{1}{2}\bar{A}^2-\frac{1}{2}\bar{A}'+\bar{E}],
    \label{eq:one}
\end{equation}
which in the case of $\bar{\alpha}=2$ and $\alpha=-1$ becomes $$ [{\Delta}V+\frac{1}{2}A^2-\frac{1}{2}A'+E]=\frac{1}{4}\bar{x}^{-2}[{\Delta}\bar{V}+\frac{1}{2}\bar{A}^2-\frac{1}{2}\bar{A}'+\bar{E}].$$
Given the expression for $A(x)$, the left side of Eq.~(\ref{eq:one}) becomes $$ax^{-2}+bx^{-1}+c+E=a\bar{x}^{-4}+b\bar{x}^{-2}+c+E.$$
where $a$,$b$, and $c$ are constants. The right side, on the other hand, becomes $$\frac{1}{4}\bar{x}^{-2}[\bar{a}\bar{x}^{-2}+\bar{b}\bar{x}^{2}+\bar{c}\bar{x}^{6}+\bar{E}]=\frac{1}{4}[\bar{a}\bar{x}^{-4}+\bar{b}+\bar{c}\bar{x}^{4}+\bar{x}^{-2}\bar{E}].$$
If $c$ and $\bar{c}$ are equal to zero, then equating these sides leads to $$E=\frac{1}{4}\bar{b},$$ $$b=\frac{1}{4}\bar{E},$$ which are exactly the relations given by the formulas $$E=-\frac{\alpha^2}{\bar{\alpha}^2}\bar{\lambda},$$ $$\lambda=-\frac{\alpha^2}{\bar{\alpha}^2}\bar{E},$$ for $b=-\lambda$ and $\bar{b}=-\bar{\lambda}$. Thus, the two Hamiltonians expressed in terms of the polynomials given above transform correctly under $x=\bar{x}^{-\frac{\bar{\alpha}}{\alpha}}$ in the case $\bar{\alpha=2}$, $\alpha=-1$.

In terms of the generators of SL(2), the Hamiltonian for the Coulomb potential (Eq.~(\ref{eq:two})) is thus given by
\begin{equation}
    H_{Cou}=C_{++}T^{++}+C_+T^+
\end{equation}
where the constants $C_+$ and $C_{++}$ are given by $$C_+=-\frac{\lambda}{6+2j},$$ $$-2C_{++}(4+k)^2-\sqrt{-2C_{++}}(4+k)=\frac{l(l+1)}{2},$$ for $k=2j-1$. Similarly, the Hamiltonian for the harmonic oscillator potential (Eq.~(\ref{eq:three})) is given by
\begin{equation}
    \bar{H}_{Osc}=\bar{C}_{-0}\bar{T}^{-0}+\bar{C}_+\bar{T}^+
    \label{eq:four}
\end{equation}
where the constants $\bar{C}_+$ and $\bar{C}_{-0}$ are given by $$\bar{C}_+(1-\bar{k}+\frac{3}{2\sqrt{-\bar{C}_{-0}}})=-\bar{\lambda},$$ $$\frac{-\bar{C}_{-0}(4+\bar{k})^2}{4}-\frac{\sqrt{-\bar{C}_{-0}(4+\bar{k})}}{2}=\frac{\bar{l}(\bar{l}+1)}{2},$$
where $\bar{k}=2\bar{j}-1$. With such definitions, the expression for the Coulomb and harmonic oscillator Hamiltonians in terms of the generators of SL(2) is made to be consistent with the transformation relating the dual potentials.

\subsection{The Spectra of the Solution}
If a Hamiltonian, $H$, is expressed in terms of the generators of SL(2), it can be possible to investigate its spectra with the ansatz wavefunction
\begin{equation}
    \psi=(1+\alpha\xi+\beta\xi^2+...+\delta\xi^{2j})e^{-a(x)}=\bar{\psi}e^{-a(x)},
\end{equation}
where $\alpha$, $\beta$, $...$ $\delta$ are constants found from the eigenvalue equation $$H\bar{\psi}=E\bar{\psi},$$
and where $$a(x)={\int}A(x)dx$$ gives the quasi-gauge transformation $e^{-a(x)}$\cite{shif98}. From Eq.~(\ref{eq:four}), and the expressions for the generators given in Eq.~(\ref{eq:five}), it is possible to quite easily solve for the spectrum of $H_{Osc}$ in the case that $C_{-0}=C_+=1$.

For $j=1$, the generators present in $H_{Osc}$ act on the ansatz as $$T^{-0}(1+\alpha\xi+\beta\xi^2)=4\beta\xi+\alpha,$$ $$T^+(1+\alpha\xi+\beta\xi^2)=2\xi+\alpha\xi^2,$$ so that the Hamiltonian leads to the eigenvalue equation
\begin{equation}
    H_{Osc}\bar{\psi}=\alpha+(4\beta+2)\xi+\alpha\xi^2=E\bar{\psi}.
\end{equation}
This expression can readily be solved, and has three eigenvalues, mirrored around zero $$E=-\sqrt{6},\alpha=-\sqrt{6},\beta=1,$$ $$E=0,\alpha=0,\beta=-2,$$ $$E=\sqrt{6},\alpha=\sqrt{6},\beta=1.$$
The spectra for $H_{Osc}$ for $j\geq1$ being found in much the same manner, with the appropriate changes to both generators and the ansatz.

Unlike the harmonic oscillator potential, it is not possible to use the same ansatz and method on the Hamiltonian for the Coulomb potential. The use of the raising operators $T^+$ to develop the $\frac{1}{r}$ potential term does not lead to a system of equations that is solvable for nonzero energies. However, it is possible to solve not for the exact $H_{Cou}$ but rather for the augmented Hamiltonian $rH_{Cou}$, with the addition of the factor of $r$ transforming the Hamiltonian into one that is exactly solvable using the properties of the SL(2) algebra\cite{shif98}. Examining Eq.~(\ref{eq:seven}), it is easy to explain why the factor of $r$ has this effect, for as $r=\bar{r}^2$, $$rH_{Cou}=\frac{1}{4}H_{Osc},$$ which is an exactly solvable Hamiltonian. The relationship between the dual potentials can thus be used to help determine the spectra of an unsolvable Hamiltonian by examining the solvable Hamiltonian of its dual partner. It is possible that this technique could be extended to potentials beyond $\frac{1}{r}$. Though $\alpha=-1,2$ are the only non-trivial partner potentials that are both expressible in terms of the generators of SL(2), $\alpha=3,4$ and $6$ are potentials which are also included within the SL(2) algebra. The Hamiltonians for these potentials suffer from the same problems as $H_{Cou}$ does, so that they are quasi-exactly solvable, rather than completely solvable. Thus, a portion of their spectra, rather than their entire spectra, can be determined in a finite number of steps from a Hamiltonian of the form Eq.~(\ref{eq:six}). It seems that it would then be possible to solve for a portion of the spectra of the $\bar{\alpha}=-6/5,-4/3$ and $-3/2$ dual potentials by utilizing the coordinate transformation \begin{equation}
    r^{-\alpha}H=\frac{\alpha^2}{\bar{\alpha}^2}\bar{H}
\end{equation}
and the quasi-solvable nature of the original Hamiltonian $\bar{H}$.

\section{\label{sec:level5}Conclusion}
Requiring that both partner potential Hamiltonians are expressible in the Lie-Algebraic construction greatly restricts the set of allowable dual potentials, removing all non-trivial potentials except for the Coulomb and Harmonic Oscillator dual potentials. These, along with the free particle potential, are the only potentials for which the transformation of coordinates that relates partner potentials is consistent with the formalism of expressing the Hamiltonians in terms of the generators of SL(2). 

The proportionality between the Coulomb Hamiltonian and the harmonic oscillator Hamiltonian allows for the spectra of the Coulomb Hamiltonian to be solved by combining the partner potential and Lie-Algebraic methods and transforming it into the exactly solvable oscillator Hamiltonian. Though the Coulomb and Oscillator potentials are the only non-trivial potentials to both be representable in terms of the generators of SL(2), other potentials can be represented alone in the SL(2) algebra, so that this method for solving the spectra could be extended to the duals of such potentials.

\appendix
\section{An Explicit Solution for $\frac{1}{r}$ and $r^2$ Potentials}
The radial Hamiltonian for the Coulomb potential is
\begin{equation}
    H=-\frac{1}{2}\frac{d^2}{dr^2}u(r)+[\frac{\lambda}{r}+\frac{l(l+1)}{2r^2}-E]u(r)=0,
    \label{eq:two}
\end{equation}
where $u(r)=r\psi(r)$ and, for simplicity, $\hbar=m=1$. The dual transformation relating the potentials $r^{-1}$ and $\bar{r}^2$ transforms the coordinates as $r=\bar{r}^2$, so that $$\frac{d^2}{dr^2}=\frac{1}{4\bar{r}^3}\frac{d}{d\bar{r}}-\frac{1}{4\bar{r}^2}\frac{d^2}{d\bar{r}^2}.$$ 
Given this, the transformation $r\rightarrow\bar{r}^2$ is able to transition between $H$, the Hamiltonian for the Coulomb $\frac{1}{r}$ potential, and $\bar{H}$, the Hamiltonian for the harmonic oscillator $\bar{r}^2$ potential. As this variable transition leads from one to the other, $H{\propto}\bar{H}$, where the proportionality is yet to be determined. Then,
$$H=\frac{1}{4\bar{r}^2}(-\frac{1}{2}\frac{d^2}{d\bar{r}^2}u(r)+\frac{1}{2\bar{r}}\frac{d}{d\bar{r}}u(r)+[\lambda+\frac{l(l+1)}{2\bar{r}^2}-E\bar{r}^2]4u(r)).$$
To simplify this, it is possible to define $$u(r)=\bar{r}^{-1/2}\bar{u}(\bar{r}),$$ so that the first derivative terms in $H$ disappear. This quasi-gauge transformation, and the definition $$V(x)={\Delta}V+\frac{1}{2}A^2(x)-\frac{1}{2}A'(x),$$ immediately leads to the Hamiltonian for the harmonic oscillator potential 
\begin{equation}
\bar{H}=-\frac{1}{2}\frac{d^2}{d\bar{r}^2}+[\bar{\lambda}\bar{r}^2+\frac{\bar{l}(\bar{l}+1)}{2\bar{r}^2}-\bar{E}]\bar{u}(\bar{r}),
\label{eq:three}
\end{equation}
where $\bar{\lambda}=-4E$, $\bar{E}=-4\lambda$, and $4l(l+1)+\frac{3}{4}=\bar{l}(\bar{l}+1)$\cite{bars2020}. Thus, under the transformation $r=\bar{r}^2$ the Coulomb Hamiltonian must be related to the oscillator Hamiltonian as
\begin{equation}
    H=\frac{1}{4\bar{r}^2}\bar{H}.
    \label{eq:seven}
\end{equation}

This result can easily be extended to arbitrary $\alpha$ and $\bar{\alpha}$, where $r=\bar{r}^{-\frac{\bar{\alpha}}{\alpha}}$. Under this transformation the second derivatives are related by $$\frac{d^2}{dr^2}=-\frac{\alpha^2}{\bar{\alpha}^2}(\frac{\bar{\alpha}}{\alpha}+1)\bar{r}^{1+2\frac{\bar{\alpha}}{\alpha}}\frac{d}{d\bar{r}}+\frac{\alpha^2}{\bar{\alpha^2}}\bar{r}^{2+2\frac{\bar{\alpha}}{\alpha}}\frac{d^2}{d\bar{r}^2}.$$
Then, again defining $$u(r)=\bar{r}^{-\frac{\bar{\alpha}}{4}}\bar{u}(\bar{r})$$ so that all of the terms with first derivatives disappear, the partner potential Hamiltonians must transform into one another as
\begin{equation}
    H=\frac{\alpha^2}{\bar{\alpha}^2}\bar{r}^{2+2\frac{\bar{\alpha}}{\alpha}}\bar{H},
\end{equation}
so that the Hamiltonians of the partner potentials are not exactly equal to one another, but rather proportional up to a power of $\bar{r}$.

\section*{Acknowledgments}

I would like to thank Professor Mikhail Shifman of the University of Minnesota for providing me with invaluable insight into the use of SL(2) generators, and for introducing me to the topic. I would also like to thank J. Rosner for looking over a draft of this paper.


\begin{thebibliography}{9}
\bibitem{newton}
Isaac Newton,
\textit{Principia Mathematica},
University of California Press, Berkeley (1934).

\bibitem{rowley}
N. Rowley,
\textit{Connection between Coulomb and oscillator problems},
J. Phys. A:Math. Gen. \textbf{12}, L7 (1979).

\bibitem{gazeau}
J. P. Gazeau,
\textit{A remarkable duality in one-particle quantum mechanics between some confining potentials and $(R+L_\epsilon^{\infty}$) potentials},
Phys. Lett. \textbf{75A}, 159 (1980).

\bibitem{bars2020}
Itzahk Bars and Jonathan L. Rosner,
\textit{Duality Between Hydrogen Atom and Oscillator Systems via Hidden SO(d,2) Symmetry and 2T-Physics},
arXiv:2001.08818 [hep-th] (23 Jan 2020).

\bibitem{johnson}
B. R. Johnson,
\textit{On a connection between radial Schrödinger equations for different power-law potentials},
J. Math. Phys. \textbf{21}, 2640 (1980).

\bibitem{ahmed}
S. A. S. Ahmed,
\textit{A transformation method of generating exact analytic solutions of the Schrödinger equation},
International Journal of Theoretical Physics. \textbf{36}, 1893 (1997).

\bibitem{turbiner}
A. Turbiner,
\textit{Quasi-exactly-solvable problems and sl (2) algebra},
Commun. Math. Phys. \textbf{118}, 467 (1988).

\bibitem{kamran}
N. Kamran and P. Olver,
\textit{Lie algebras of differential operators and Lie-algebraic potentials},
J. Math. Anal. Appl. \textbf{145}, 342 (1990).

\bibitem{ushveridze}
A. Ushveridze,
\textit{Quasi-exactly solvable problems in quantum mechanics},
IOP Publishing, Bristol (1994).

\bibitem{shif98}
M. Shifman,
\textit{Quasi-Exactly Solvable Spectral Problems},
ITEP Lectures in Particle Physics and Field Theory. \textbf{2}, 775 (1998).

\end{thebibliography}
\end{document}